\documentclass[twocolumn]{aastex631}
\usepackage{amsmath}
\usepackage{float}
\usepackage{breqn}
\usepackage{multirow}

\newcommand{\planet}{{Kepler-167~e}}

\accepted{in ApJL on November 1, 2022}

\shorttitle{A 16 Hour Transit Observed by the Unistellar Network}
\shortauthors{Perrocheau et al.}

\begin{document}

\title{A 16 Hour Transit of Kepler-167 e Observed by the Ground-based Unistellar Telescope Network}

\correspondingauthor{Tom Esposito}
\email{tesposito@seti.org}

\author[0000-0003-0538-8561]{Amaury Perrocheau}
\affiliation{SETI Institute, Carl Sagan Center, 339 Bernardo Ave, Suite 200, Mountain View, CA 94043, USA}
\affiliation{Unistellar, 5 all\'ee Marcel Leclerc, b\^{a}timent B, Marseille, 13008, France}

\author[0000-0002-0792-3719]{Thomas M. Esposito}
\affiliation{SETI Institute, Carl Sagan Center, 339 Bernardo Ave, Suite 200, Mountain View, CA 94043, USA}
\affiliation{Unistellar, 5 all\'ee Marcel Leclerc, b\^{a}timent B, Marseille, 13008, France}
\affiliation{Astronomy Department, University of California, Berkeley, CA 94720, USA}

\author[0000-0002-4297-5506]{Paul A.\ Dalba}
\altaffiliation{Heising-Simons 51 Pegasi b Postdoctoral Fellow}
\affiliation{SETI Institute, Carl Sagan Center, 339 Bernardo Ave, Suite 200, Mountain View, CA 94043, USA}
\affiliation{Department of Astronomy and Astrophysics, University of California, Santa Cruz, CA 95064, USA}

\author[0000-0001-7016-7277]{Franck Marchis}
\affiliation{SETI Institute, Carl Sagan Center, 339 Bernardo Ave, Suite 200, Mountain View, CA 94043, USA}
\affiliation{Unistellar, 5 all\'ee Marcel Leclerc, b\^{a}timent B, Marseille, 13008, France}

\author[0000-0001-7801-7425]{Arin M. Avsar}
\affiliation{Lunar and Planetary Laboratory, University of Arizona, Tucson, AZ 85721, USA}


\author{Ero Carrera}
\affiliation{Unistellar Citizen Scientist}

\author{Michel Douezy}
\affiliation{Unistellar Citizen Scientist}

\author{Keiichi Fukui}
\affiliation{Unistellar Citizen Scientist}

\author{Ryan Gamurot}
\affiliation{Unistellar Citizen Scientist}

\author{Tateki Goto}
\affiliation{Unistellar Citizen Scientist}

\author[0000-0003-4091-0247]{Bruno Guillet}
\affiliation{Unistellar Citizen Scientist}

\author{Petri Kuossari}
\affiliation{Unistellar Citizen Scientist}

\author{Jean-Marie Laugier}
\affiliation{Unistellar Citizen Scientist}

\author[0000-0003-0828-6368]{Pablo Lewin}
\affiliation{The Maury Lewin Astronomical Observatory, Glendora, CA, USA}
\affiliation{Exoplanet Watch Citizen Scientist}

\author{Margaret A. Loose}
\affiliation{Unistellar Citizen Scientist}

\author{Laurent Manganese}
\affiliation{Unistellar Citizen Scientist}

\author{Benjamin Mirwald}
\affiliation{Unistellar Citizen Scientist}

\author{Hubert Mountz}
\affiliation{Unistellar Citizen Scientist}

\author{Marti Mountz}
\affiliation{Unistellar Citizen Scientist}

\author{Cory Ostrem}
\affiliation{Unistellar Citizen Scientist}

\author{Bruce Parker}
\affiliation{Unistellar Citizen Scientist}

\author{Patrick Picard}
\affiliation{Unistellar Citizen Scientist}

\author{Michael Primm}
\affiliation{Unistellar Citizen Scientist}

\author{Justus Randolph}
\affiliation{Unistellar Citizen Scientist}

\author{Jay Runge}
\affiliation{Unistellar Citizen Scientist}

\author{Robert Savonnet}
\affiliation{Unistellar Citizen Scientist}

\author[0000-0002-6250-5608]{Chelsea E. Sharon}
\affiliation{Unistellar Citizen Scientist}
\affiliation{Yale-NUS College, 16 College Ave West 01-220, 138527, Singapore}

\author{Jenny Shih}
\affiliation{Unistellar Citizen Scientist}

\author{Masao Shimizu}
\affiliation{Unistellar Citizen Scientist}

\author{George Silvis}
\affiliation{Exoplanet Watch Citizen Scientist}

\author{Georges Simard}
\affiliation{Unistellar Citizen Scientist}

\author{Alan Simpson}
\affiliation{Unistellar Citizen Scientist}

\author{Thusheeta Sivayogan}
\affiliation{Unistellar Citizen Scientist}

\author{Meyer Stein}
\affiliation{Unistellar Citizen Scientist}

\author{Denis Trudel}
\affiliation{Unistellar Citizen Scientist}

\author{Hiroaki Tsuchiyama}
\affiliation{Unistellar Citizen Scientist}

\author[0000-0002-4309-6343]{Kevin Wagner}
\altaffiliation{NASA Hubble Fellowship Program – Sagan Fellow}
\affiliation{Unistellar Citizen Scientist}
\affiliation{Department of Astronomy and Steward Observatory, University of Arizona, 933 N Cherry Ave, Tucson, AZ 85721, USA}

\author{Stefan Will}
\affiliation{Unistellar Citizen Scientist}



\begin{abstract}

More than 5,000 exoplanets have been confirmed and among them almost 4,000 were discovered by the transit method. However, few transiting exoplanets have an orbital period greater than 100 days. Here we report a transit detection of \planet, a ``Jupiter analog'' exoplanet orbiting a K4 star with a period of 1,071 days, using the Unistellar ground-based telescope network. From 2021 November 18 to 20, citizen astronomers located in nine different countries gathered 43 observations, covering the 16 hour long transit. Using a nested sampling approach to combine and fit the observations, we detected the mid-transit time to be UTC 2021 November 19 17:20:51 with a 1$\sigma$ uncertainty of 9.8 minutes, making it the longest-period planet to ever have its transit detected from the ground. This is the fourth transit detection of \planet, but the first made from the ground. This timing measurement refines the orbit and keeps the ephemeris up to date without requiring space telescopes. Observations like this demonstrate the capabilities of coordinated networks of small telescopes to identify and characterize planets with long orbital periods.

\end{abstract}



\section{Introduction} \label{sec:intro}

Around 800 exoplanets are known to have an orbital period of more than 100 days, based on data extracted from the NASA Exoplanet Archive\footnote{As of 2022 September 2.} (NEA; \citealt{Akeson_2013, nea}). Out of these 800, only ${\sim}170$ have been detected using the transit method. This is primarily because the probability of observing an exoplanet transit decreases as the planet's orbital semi-major axis increases, but also because the relatively small fraction of its orbit that a planet spends in transit requires almost continuous observation for detection \citep{Beatty_2008}. While the probability of detection does increase with orbital eccentricity, with periastron becoming smaller, the time between transits remains long \citep[e.g.,][]{Kane_2007,Dalba_2021}. Freed from the constraints of ground-based observations, the \textit{Kepler} \citep{Borucki_2010} and \textit{Transiting Exoplanet Survey Satellite} \citep[TESS;][]{Ricker_2014} missions enabled great progress in the search for long period exoplanets by providing long continuous baseline transit surveys (\citealt{Wang_2015}, \citealt{Uehara_2016}, \citealt{Herman_2019}), although for TESS this is true only near the ecliptic poles.

Studying these long period planets is crucial to understanding our own solar system. Jupiter has an orbital period of 12 years but dominates the planets' gravitational effects and is partly responsible for the current architecture of the solar system due to its migrations (e.g., \citealt{Tsiganis2005}, \citealt{Walsh_2011}). It is possible to observe some of these long period exoplanets transiting only once every few years and it requires substantial observation time on space telescopes to detect those with periods that are unknown or poorly constrained \citep{Foreman-Mackey_2016}. Observing them from the ground is even harder because of the Sun's daily interference and the long durations of these transits; of the confirmed planets with period $>100$ days and a measured duration, all but one have a duration $>3.5$ hours (NEA), meaning at least a 7 hour observation is required to gather equal amounts of in-transit and out-of-transit data. From the ground, therefore, only a network of telescopes spanning the globe is capable of capturing enough of the transit to lead to a detection \citep{VonBraun2009}. The longest-period transiting planet previously detected was HIP~41378~f (542 days) and it required four professional telescopes in Chile, Spain and Italy, only two of which actually captured a combined ${\sim}3$ hours of in-transit data out of the 19 hour event \citep{Bryant_2021}. Similarly, two observing sites in Spain \citep{Garcia-Melendo2009} and the United Kingdom \citep{Fossey2009} detected only egress of the first observed 12-hour-long transit by HD~80606~b (111 day period); admirable considering the mid-transit date was only predicted within a ${\sim}1$ day window \citep{Laughlin2009}. Re-detection of a nearly complete transit by HD~80606~b by \citet{Pearson_2022} via global coordination of  ground-based telescopes, including Unistellar Network eVscopes, illustrates the value of geographically distributed observations.

Here, we present the first ground-based detection of the transiting ``Jupiter analog'' \planet\ with observations by the Unistellar Network \citep{marchis_2019}. \planet\  has  a period of $1071.23205 \substack{+0.00059 \\ -0.00058}$ days \citep{Chachan_2022} and transit duration of $16.25 \pm 0.07 $ hours (taken as the mean duration of 16,000 sample models drawn from the Chachan et al. 2022 posterior distribution). With a radius between that of Jupiter and Saturn (0.9 $R_J$), it has about the same mass as Jupiter (1.01 $M_J$) and a low orbital eccentricity (0.06; \citealt{Kipping_2016}). It orbits a K4V star with $V$=14.284, along with three inner super-Earth planets that also transit \citep{Kipping_2016}. The \textit{Kepler} mission enabled the first two observations, in 2010 and in 2013 \citep{Kipping_2016}, and half a transit was observed in 2018 with the \textit{Spitzer Space Telescope} \citep{Werner_2004} to rule out transit timing variations (TTV) of at least 34 minutes to 3$\sigma$ confidence \citep{Dalba_2019}. Thanks to these works, the uncertainty on the mid-transit time ($T_0$) of the 2021 transit prediction was just $\sigma_{T_0}=3.0$~min.

The Unistellar Network is composed of ${\sim}10,000$ digital telescopes called Enhanced Vision Telescopes, or ``eVscopes''  \citep{Marchis_2020}. Citizen astronomers in possession of an eVscope can be found all over the world. These telescopes have already been used to observe exoplanet transits (e.g., HD~80606~b by \citealt{Pearson_2022}, WASP-148~b by \citealt{Wang_2022}, TOI-2180~b by \citealt{Dalba_2022}), time asteroid occultations \citep{Cazaneuve_2022}, and monitor the James Webb Space Telescope (JWST) \citep{Lambert_2022}.

We describe here our observations and analyses that led to the detection of the \planet \ transit. In Section \ref{sec:observation}, we present the photometric data gathered from the Unistellar Network. Section \ref{sec:Methods} describes the methods for combining the observations and modeling the transit. Section \ref{sec:result} presents the results and finalized light curve. Finally, Section \ref{sec : discussion} discusses the validity of our results and relevance of a ground-based telescope network for observing long duration exoplanet transits.

\section{Observations}
\label{sec:observation}

\subsection{Unistellar eVscopes}

 The eVscopes are portable digital consumer telescopes manufactured and distributed by Unistellar. Two hardware models are currently in use: the eVscope 1 (same as the eQuinox model) and the eVscope 2. They both have an aperture of 114.3 mm and a focal length of 450 mm. Both also use off-the-shelf Sony CMOS sensors located at the prime focus with low read noise and dark current, with a Bayer filter. The IMX224 sensor in the eVscope 1 has a field of view (FOV) of $37\arcmin \times 28\arcmin $ with a pixel scale of 1$\farcs$72 per pixel and the IMX347 in the eVscope 2 has an FOV of $45\arcmin \times 34\arcmin$ with a pixel scale of 1$\farcs$33 per pixel. Each eVscope is identified by a 6 digit alphanumerical string that is abbreviated in this paper to the first three characters. A suffix ``-B'' is added when the citizen scientist observed for a second time during the campaign with the same telescope.

\subsection{Observation Campaign}

\citet{Dalba_2019} predicted $T_0$ of this 2021 transit to be 2459538.2189 $\text{BJD}_{\text{TDB}} \, \pm 0.0021$ (a 3.0 min uncertainty). They also ruled out TTV's greater than 34 min to 3$\sigma$ significance. Using this information, the SETI Institute, scientific partner of Unistellar, made a call to the Unistellar Network on 2021 November 4 to observe the upcoming transit. In response, 31 citizen astronomers and two professional astronomers from our team observed the target star between 2021 November 18 and 20 using Unistellar eVscopes, producing 43 individual data sets. The exposure time was set at 3.970 s and the gain at 0.02245 e- ADU$^{-1}$ (40 dB). Detailed information about the observations is given in Table \ref{table:observation table}. To have enough continuous dark time to record the $>$16 hour transit duration (temporarily ignoring the time also required to record important out-of-transit baseline data), a single observing site would need an extreme latitude of ${>}76\degr$ N; thus the necessity to combine observations from multiple longitudes around the world. Two citizen astronomers not affiliated with the Unistellar Network also observed Kepler-167 during this campaign, discussed in Section \ref{sec:EW_obs}.

\begin{figure}[h!]
    \centering
    \includegraphics[width=8cm]{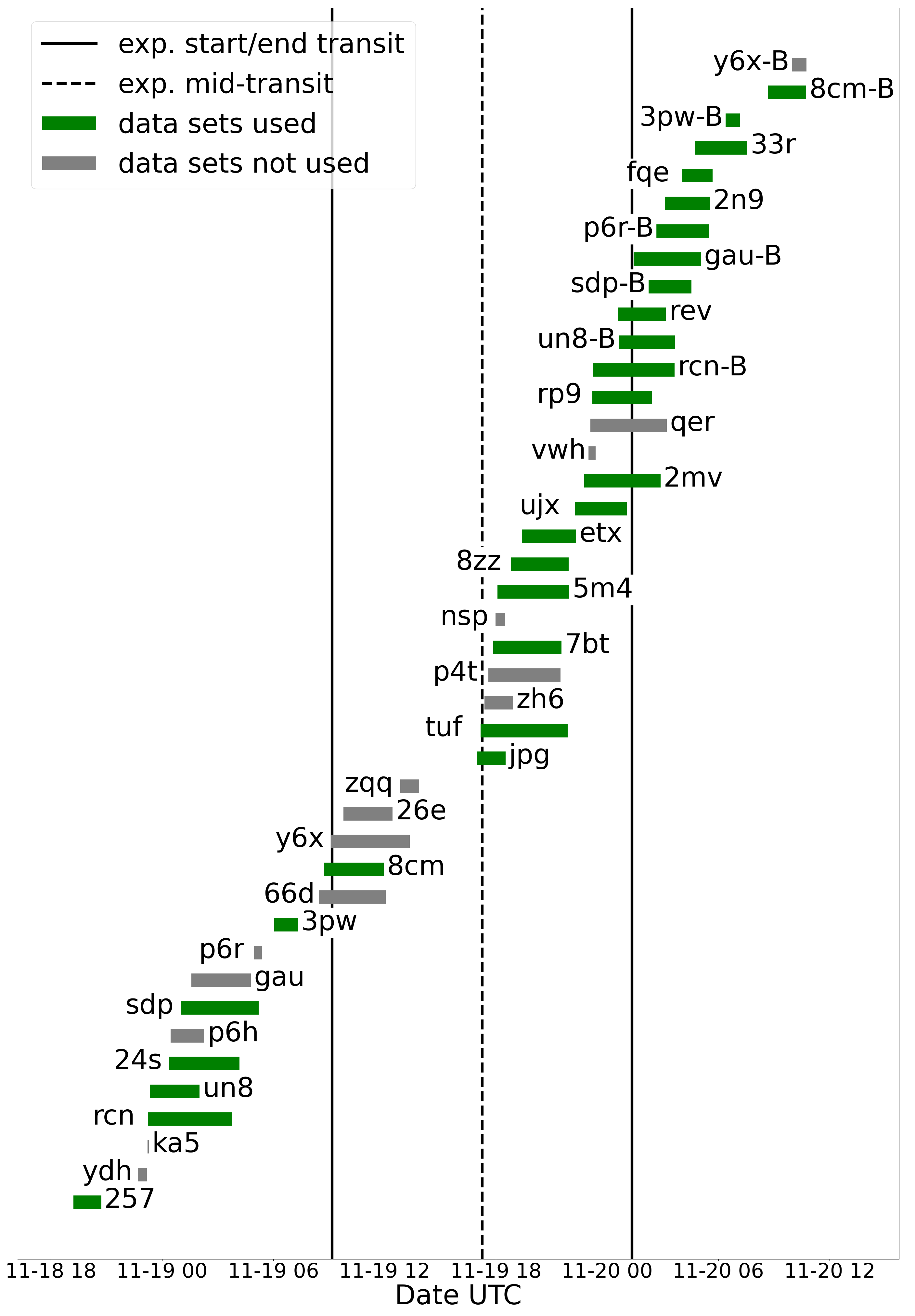}
    \caption{Time frames for eVscope observations of \planet. 27 out of the 43 observations (green) were used to construct the final light curve. Vertical lines mark the predicted transit start, middle, and end times.}
    \label{fig: timeframe}
\end{figure}

\subsection{Observation Selection}

The observations were made from various sites spread around Earth, leading to variable data quality. At that time, the Moon was ${\sim}$100\% illuminated and located ${\sim}$103$^{\circ}$ from the target. Some data were not usable because of poor weather, poor atmospheric seeing, or sub-optimal calibration of the telescope focus or collimation. Moreover, the relatively high sensor gain used to observe this $V$=14.3 target caused an unexpectedly large non-linear increase in the background sky brightness that saturated some observations made from light-polluted areas. To avoid these data, we excluded observations where at least one pixel in the target star's point-spread function was saturated or the scatter in the relative flux residuals after model subtraction was $>10$\%. As a result, only 27 out of the 43 data sets were used in the analysis. The time distribution of observations can be found in Figure \ref{fig: timeframe} in relation to the predicted transit times.

\section{Methods} \label{sec:Methods}

\subsection{Photometry}

The raw images received from the eVscopes in FITS format are first dark subtracted, plate solved using \texttt{Astrometry.net} \citep{Lang_2010}, aligned, and then stacked in groups of 30 representing 119.1~s of integration time. This image stacking increases star signal-to-noise ratio (SNR) by reducing random noise and further mitigates photometric noise by averaging over the differing spectral responses of individual pixels caused by the Bayer filter \citep[e.g.,][]{Lambert_2022}.

Next we performed differential aperture photometry on the stacked images using the \texttt{Photutils} Python package \citep{larry_bradley_2020}. This allowed us to locate every star in an image and measure its position, flux, and uncertainty on that flux. For each data set, we identified several potential comparison stars that satisfied our criteria of being located within $3 \arcmin$ of the target star, unsaturated in the individual images, and not blended with other sources. For consistency between data sets, we chose to use the same single comparison star to compute all of our relative fluxes: Gaia DR3 2051932862036924416 ($19^{\text{h}}30^{\text{m}} 40\fs 31$ +$38\text{°} 19\arcmin 34 \farcs 14$), slightly fainter than Kepler-167 and located $1\farcm25$ away. Being similar in magnitude to the target star meant this comparison star was also the highest SNR option that was unsaturated. We then optimized the sizes of the circular flux apertures and annular background apertures for both target and comparison star independently by minimizing the scatter in the individual star fluxes within each data set.

\subsection{Detrending}

With the photometry done, we began assembling the light curve of the complete event from the individual light curves of each data set. By taking the relative flux ratio between the target and the comparison star, long term trends in the fluxes and image-to-image photometric variations were largely removed. Trends due to differential atmospheric extinction (i.e., airmass), in particular, were greatly mitigated by the nearly identical $G_{BP}-G_{RP}$ colors of the target and comparison stars, given by the \textit{Gaia} DR3 catalog \citep{GAIA_2016, GAIA_2022} as $(G_{BP}-G_{RP})_{targ}=1.227932$ and $(G_{BP}-G_{RP})_{comp}=1.190314$). No other detrending was applied. Finally, the relative fluxes of each data set are divided by their medians to normalize them to approximately 1.0 in preparation for further normalization during modeling.

We found most data sets' relative fluxes to be approximately flat over time but some did contain significant slopes. The timing of the most steeply sloped fluxes coincided with the transit's expected ingress and egress times, suggesting they were real physical features; however, we did not assume this to be the case during our analysis.

\begin{table*}[ht!]
\caption{Transit Parameter Values} \label{tab:transit parameters}
\centering
\begin{tabular}{c c c c}
\tablewidth{0pt}
\hline
\hline
Parameter  & Expected Value & Distribution Priors  & Best Fit 
\\
 \hline
$T_0$ ($\text{BJD}_{\text{TDB}}$)  & $2459538.2189 \pm 0.0021$  & $\mathcal{U}[2459538.1793 , 2459538.2585]$ & $2459538.2225 \pm 0.0068$ \\
$D$ (\%) & $1.540_{\, -0.032}^{+0.027}$  & $\mathcal{N}(1.540,4)$ &  $1.77\pm 0.42$ \\
$W \times P$ (minutes)  & $975 \pm 4$  \\
$P$ (days)  & $1071.23205_{\, -0.00058}^{+0.00059}$  & \multicolumn2c{\multirow{4}{8em}{ \centering Fixed}}   \\
$b$  & $0.271 _{\, -0.073}^{+0.051}$   \\
$e$  & 0 \\
$h_1$ &  $ 0.684 \pm  0.011 $     \\
$h_2$   & $0.434  \pm    0.050 $     \\
$c_1$ (257) & 1 &  &  $0.997 \pm 0.007 $ \\  
$c_2$ (rcn) & 1 &  & $1.001 \pm 0.003 $ \\  
$c_3$ (un8) & 1 &  & $ 0.998 \pm 0.006 $ \\  
$c_4$ (24s) & 1 &  & $ 0.990 \pm 0.006 $ \\  
$c_5$ (sdp) & 1 &  & $0.992 \pm 0.015 $ \\  
$c_6$ (3pw) & 1 &  & $ 1.001 \pm 0.007 $ \\  
$c_7$ (8cm) & 0.993 &  & $ 0.992 \pm 0.004$ \\  
$c_8$ (jpg) & 0.978 &  & $ 0.982 \pm 0.009 $ \\  
$c_9$ (tuf) & 0.978 &  & $ 0.978 \pm 0.006 $ \\  
$c_{10}$ (7bt) & 0.979 &  & $0.982 \pm 0.009$ \\  
$c_{11}$ (5m4) & 0.979 &  & $ 0.987\pm 0.008 $ \\  
$c_{12}$ (8zz) & 0.979 & $\mathcal{N}(1,0.0004)$ & $ 0.987 \pm 0.007 $ \\  
$c_{13}$ (etx) & 0.979 &  & $ 0.980 \pm 0.015 $ \\  
$c_{14}$ (ujx) & 0.988 &  & $ 0.991 \pm 0.005$ \\  
$c_{15}$ (2mv) & 0.993 &  & $0.993 \pm 0.004 $ \\  
$c_{16}$ (rp9) & 0.993 &  & $ 0.989 \pm 0.005 $ \\  
$c_{17}$ (rcn-B) & 0.996 &  & $ 0.994\pm 0.004$ \\  
$c_{18}$ (un8-B) & 0.996 &  & $0.998 \pm 0.004$ \\  
$c_{19}$ (rev) & 0.999 &  & $0.993 \pm 0.008$ \\  
$c_{20}$ (sdp-B) & 1 &  & $ 0.998 \pm 0.008 $ \\  
$c_{21}$ (gau-B) & 1 &  & $ 0.998 \pm0.005 $ \\  
$c_{22}$ (p6r-B) & 1 &  & $ 0.998\pm 0.003$ \\
$c_{23}$ (2n9) & 1 &  & $ 0.987 \pm 0.011 $ \\  
$c_{24}$ (fqe) & 1 &  & $ 0.997\pm 0.009$ \\  
$c_{25}$ (33r) & 1 &  & $1.003\pm 0.006$ \\  
$c_{26}$ (3pw-B) & 1 &  & $0.987 \pm 0.008$ \\  
$c_{27}$ (8cm-B) & 1 &  & $0.998 \pm 0.004 $ \\
 \hline
 
\end{tabular}

\tablecomments{The transit duration is $W \times P$. Parameters $h_1$ and $h_2$ are stellar limb darkening coefficients within the \texttt{PYCHEOPS} model \citep{Maxted_2022}. The normal prior on the normalization coefficients $c_i$ has a mean of 1 and a standard deviation of 2\%. The expected value of each $c_i$ is the mean of the relative flux during the times of observation $i$ as measured from a model built from the ``expected'' values for all parameters. The distribution priors are written as $\mathcal{U}[a,b]$ and $\mathcal{N}(\mu, \sigma^2)$ for the uniform and normal distributions, respectively.}

\end{table*}

\subsection{Light Curve Normalization and Modeling}

Extracting the transit's signal and timing from these multiple data sets requires combining them into one light curve and comparing that with a model. Differential photometry does not establish the absolute brightness of the target star, thus in order to properly construct a combined light curve we need to normalize the individual light curves relative to some reference baseline. We start with the assumption that relative fluxes measured outside of the transit will have a mean of 1.0, which becomes our reference baseline.

To normalize fluxes within our model, we multiply a scalar coefficient $c_i$ with the relative fluxes $y_{i,j}$ of every individual light curve $i$, where $j$ denotes the time of each measurement. Thus, every individual light curve is free to move up and down relative to the reference baseline. We then construct a transit light curve model using the \texttt{PYCHEOPS} package \citep{Maxted_2022} and fit it to the combination of all normalized fluxes. Stellar limb darkening effects are taken into account using the power-2 law as implemented in the \texttt{qpower2} algorithm  \citep{Maxted_2019}.

The parameters used to compute the transit model are given in Table \ref{tab:transit parameters}. In our approach, which aims to retrieve the transit times and depth assuming the other transit properties are well known, we fix the following model parameters to the latest values published by \citet{Chachan_2022}: the planet's orbital period $P$, transit duration (parameterized as $P \times W$ where $W$ is the length of the transit in orbital phase units), impact parameter $b$, eccentricity $e$ (assumed to be 0), and stellar limb darkening coefficients $h_1$ and $h_2$. As a result, only the mid-transit time $T_0$ and the transit depth $D$ can vary. The eccentricity measurements from \citet{Kipping_2016} and \citet{Chachan_2022} are both consistent with zero, so we assume the simplest case of a circular orbit. The limb darkening coefficients $h_1$ and $h_2$ are transformed parameters from the power-2 law described in \citet{Maxted_2022}.

\subsection{Dynamic Nested Sampling}
\label{Dynamic nested sampling}

To fit our model to the data we used Bayesian statistics via a dynamic nested sampling approach \citep{Higson_2019}. This method lets us estimate parameters from a highly multidimensional model. The sampling algorithm was implemented with the \texttt{dynesty} Python package \citep{Speagle_2020}, which required two functions: the prior transformation and the likelihood function.

The prior transformation converts a hypercube of values between 0 and 1 to physical scales. To avoid biasing the result toward a particular mid-transit time, we used a uniform prior for $T_0$. The prior bounds are chosen to be the expected $T_0$ $\pm$ the upper limit of the smallest TTV magnitude ruled out at 5$\sigma$ significance by \cite{Dalba_2019}, i.e., $\pm 57$ min. For $D$ and normalization coefficients $c_i$, we chose normal priors with the expected value as the mean and a 2\% standard deviation. The values chosen are listed in Table \ref{tab:transit parameters}. These priors were specifically chosen to be loose so the intrinsic likelihood would primarily drive the fit.

We assume that the flux measurements are independent and therefore drawn from a normal distribution. We consider this assumption to be justified because we confirmed that the time-averaged root mean squared of the full light curve's binned flux residuals (using the best-fit model) decreased with bin size at the rate that theory predicts for residuals dominated by uncorrelated white noise \citep{Cubillos2017}. The overall likelihood is then obtained by multiplying together the individual measurement likelihoods. Finally, because it is easier in practice to maximize the logarithm of the likelihood, we are left with an effective likelihood function of:

\begin{equation}
    \log \mathcal{L}=-\frac{1}{2}\sum_{i}\sum_j \frac{\big(c_iy_{i,j}-f_j\big)^2}{\sigma_{i,j}^2}
\end{equation}

where $\sigma_{i,j}$ are the uncertainties associated with each relative flux measurement $y_{i,j}$, and $f_j$ are the modeled light curve fluxes at measurement time $j$.

We consider that the model fitting procedure has converged when $\log \mathcal{Z} < 0.1 $, where $\mathcal{Z}$ is the Bayesian model evidence.

\begin{figure*}[ht!]
    \centering
    \includegraphics[width=18cm]{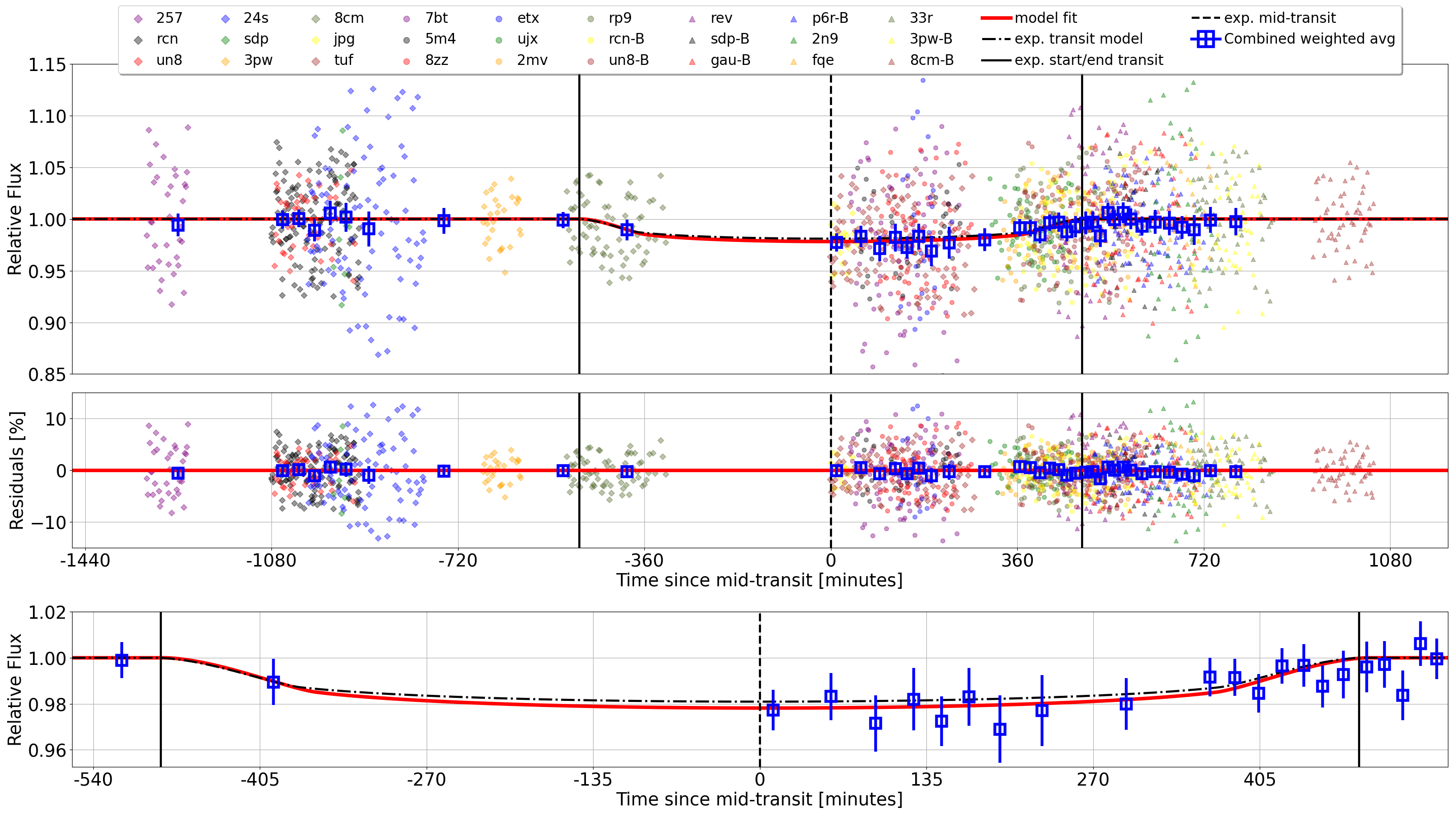}
    \caption{Transit light curve of \planet\ (top) and residuals (middle) as observed by the Unistellar Network. Relative fluxes from individual data sets are plotted with different colors and symbols. The maximum likelihood model corresponding to the values in Table \ref{tab:transit parameters} is the red line and the time axis is defined as time since that model's mid-transit time. The blue squares are weighted average fluxes of the combined data. The expected transit model and its transit start, middle, and end times are plotted as black broken lines. The bottom panel is the same as the top panel but zooms in on the times the transit occurred and only shows the weighted average fluxes.}
    \label{fig:Light curve}
\end{figure*}

\section{Results} \label{sec:result}

Using the nested sampling approach described in Section \ref{sec:Methods}, we obtained the maximum likelihood (``best-fit'') light curve at the top of Figure \ref{fig:Light curve}, along with an estimation of the transit parameters. We measure $T_0$ to be UTC 2021 November 19 17:20:51 $\pm$ 9.8 min ($2459538.2225 \pm 0.0068$ $\text{BJD}_{\text{TDB}}$) within the 68\% confidence interval. This is statistically consistent with the predicted mid-transit time of UTC 2021 November 19 17:15:35 $\pm$ 3.0 min ($2459538.2189 \pm 0.0021$ $\text{BJD}_{\text{TDB}}$). The depth of the transit is measured to be $D=1.77 \pm 0.42$\% compared to the predicted $1.540^{+0.027}_{\,-0.032}$\%. The computed values for the normalization coefficients $c_i$ are listed by data set in Table \ref{tab:transit parameters}.

To test that our normalization of the individual light curves is not biasing our timing measurements, we also fit a transit model to a light curve that excluded the fully in-transit data, which involved removing fluxes from the following data sets: \textit{jpg, tuf, 7bt, 5m4, 8zz} and \textit{etx}. This left only data sets containing at least some out-of-transit data (based on our maximum likelihood model). The resulting $T_0$ computed without in-transit data is UTC 2021 November 19 17:07:45 $\pm$ 11 min ($2459538.2134 \pm 0.0073$ $\text{BJD}_{\text{TDB}}$) and the depth is $1.20 \pm 0.48$\%. This transit depth is still consistent with the prediction and the two configurations --- with and without the in-transit data --- are consistent with each other at the $1\sigma$ level.

\section{Discussion} \label{sec : discussion}

\subsection{Exoplanet Watch Observation} \label{sec:EW_obs}
\begin{figure}[h!]
    \centering
    \includegraphics[width=\columnwidth]{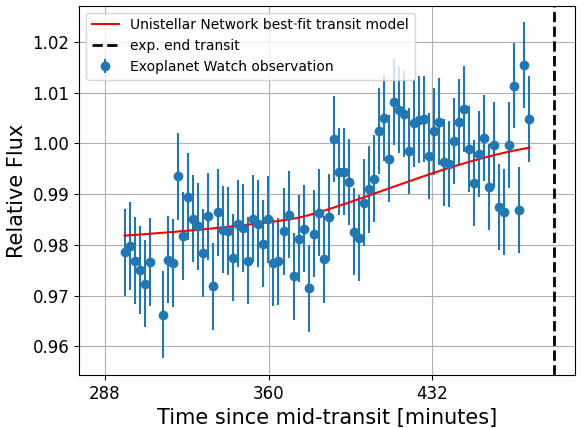}
    \caption{Exoplanet Watch observation of the \planet\ egress. Relative fluxes are the blue dots, with corresponding $1\sigma$ error bars. The best-fit transit model from fitting only Unistellar Network data is overlaid in red. This model was not fit to the Exoplanet Watch observations.}
    \label{fig:Exoplanet Watch}
\end{figure}

Two citizen astronomers (Lewin and Silvis) participating in NASA's Exoplanet Watch program (EW; \citealt{Zellem_2020}) observed Kepler-167 during the same campaign in November 2021. We acquired their data from the public database hosted by the American Association of Variable Star Observers\footnote{\url{https://www.aavso.org/}}. Lewin observed a flat light curve in data taken well before the expected transit ingress, but Silvis observed during the expected egress with a 12 inch (30 cm) telescope (Table~\ref{table:observation table}). Those relative fluxes, after being detrended with a standard airmass correction function, are plotted in Figure \ref{fig:Exoplanet Watch} and overlaid with the maximum likelihood transit model computed from our ``best-fit'' parameters of Table \ref{tab:transit parameters}. The systematic ``wiggle'' in the light curve at $X\approx388$--407 corresponds to a temporary increase in image background level from environmental light pollution (a house light being turned on). Following this, increasing airmass (from 1.4 to 1.6) and a broadening PSF may have introduced additional systematic noise via imperfect background subtraction. The last ten flux points were strongly affected by the local horizon vignetting the entrance pupil; a systematic error not encompassed by their error bars.

Without conducting any additional fitting, the $\chi^2$ and the reduced $\chi^2_\nu$ values between our transit model and the EW fluxes are $\chi^2=65.9$ and $\chi_\nu^2=0.82$, respectively. In contrast, the corresponding values when comparing the EW fluxes with a flat model normalized to the median of the detrended data are $\chi^2\text{(flat)}=146.6 \text{ and } \chi_\nu^2\text{(flat)}=1.81$. We consider this an independent confirmation that the Unistellar Network data successfully detected the egress of the transit.

\subsection{Observing Exoplanet Transits with a Network of Small Ground-Based Telescopes}

This successful measurement of \planet\ proves that it is possible to detect a long period, long duration exoplanet transit with a network of small ground-based telescopes. Even though we knew the exoplanet parameters precisely before the transit occurred in this case, we can reproduce this result for planet candidates that have their properties less constrained. Using small telescopes such as the eVscope therefore permits the detection and confirmation of planet candidates with imprecisely known orbital periods without requiring observation time on in-demand space telescopes such as JWST or the \textit{Hubble Space Telescope} (HST). A mission like ESA's current CHaracterising ExOPlanet Satellite (CHEOPS), which can do long duration follow-up, could also benefit from a small-telescope network because it has its own observability limitations owing to its orbit and Earth occultations \citep{Benz_2020}.

Keeping transiting planet ephemerides ``fresh'' (i.e the 1$\sigma$ uncertainty in the mid-transit time to less than half the transit duration) is crucial; for instance, most of the ephemerides coming from \textit{TESS} will be outdated only one year after the last observation \citep{Dragomir_2020}. This concern is not limited to transiting planets, as radial velocity-detected planets also need to be monitored \citep{Kane2009}. \citet{Zellem_2020} proved that uncertainties of just 1 minute in a planet's orbital period and mid-transit time will produce an uncertainty of ${\sim}15$ hours in that planet's mid-transit time ten years later. Thus, regular re-detections with small ground-based telescopes are important to paving the way for detailed spectroscopic characterization by JWST and subsequent generations of exoplanet characterizing facilities.

\subsection{Transit Timing Consistency and Future Studies}

Our timing measurement supports the previous finding by \citet{Dalba_2019} that Kepler-167~e's ephemeris does not contain large TTV's. The difference between our measured mid-transit epoch and the initial transit epoch recomputed by \citet{Chachan_2022} is $4284.9349 \pm 0.0068$ days, which when divided by four equates to an orbital period of $1071.2337 \pm 0.0017$ days. This is in line with the \citet{Kipping_2016} value of $P=1071.23228 \pm 0.00056$ days and $P=1071.23205_{\, -0.00058}^{+0.00059}$ from \citet{Chachan_2022}, though about 2 min longer than both. Using our best-fit mid-transit time as a reference epoch to compute future mid-transit times, we obtain values that fall within the \citet{Dalba_2019} 1$\sigma$ predictions, but we do not improve upon their precision so we do not report new values here. The relatively large uncertainty on the previous epoch's measurement from \textit{Spitzer} data also prevents us from tightly constraining TTV's over the planet's last orbit, but we can rule out 24 and 73 min TTV's to $1\sigma$ and $3\sigma$ significance, respectively. Our own measurement uncertainty will be the dominant factor in ruling out even smaller TTV's over the next orbit, thus improved timing precision for the next transit (e.g., with greater telescope coverage from our network) will be a future objective.

The lack of large TTV's means that future observations by telescopes like HST or JWST will not run the risk of missing substantial portions of valuable in-transit time for \planet. It is rare for such a long period (and low temperature) planet to be so well characterized with an unambiguous ephemeris and measured mass and radius. Therefore, we encourage additional work to explore this planet's suitability for expanded characterization of its atmosphere and potential exomoons.

\clearpage 

\section{Acknowledgements}

The authors thank all the citizen astronomers that participated in this observation campaign and contributed their precious data. We also thank Rob Zellem for his contributions to this research effort and the anonymous reviewer for their thoughtful feedback.

This research has made use of the NASA Exoplanet Archive, which is operated by the California Institute of Technology, under contract with the National Aeronautics and Space Administration under the Exoplanet Exploration Program. This work has also made use of data from the European Space Agency (ESA) mission {\it Gaia} (\url{https://www.cosmos.esa.int/gaia}), processed by the {\it Gaia} Data Processing and Analysis Consortium (DPAC,\url{https://www.cosmos.esa.int/web/gaia/dpac/consortium}). Funding for the DPAC has been provided by national institutions, in particular the institutions participating in the {\it Gaia} Multilateral Agreement.

This publication makes use of data products from Exoplanet Watch, a citizen science project managed by NASA's Jet Propulsion Laboratory on behalf of NASA's Universe of Learning. This work is supported by NASA under award number NNX16AC65A to the Space Telescope Science Institute, in partnership with Caltech/IPAC, Center for Astrophysics|Harvard \& Smithsonian, and NASA Jet Propulsion Laboratory. We acknowledge with thanks the variable star observations from the AAVSO International Database contributed by observers worldwide and used in this research.


T.M.E., F.M., and the Unistellar Network's science campaigns are supported in part by a grant from the Gordon and Betty Moore Foundation. P.D. acknowledges support from a National Science Foundation (NSF) Astronomy and Astrophysics Postdoctoral Fellowship under award AST-1903811 and a 51 Pegasi b Postdoctoral Fellowship from the Heising-Simons Foundation.

\software{Astrometry.net \citep{Lang_2010}, Astropy \citep{Astropy_2013, Astropy_2018}, dynesty \citep{Speagle_2020}, Photutils \citep{larry_bradley_2020}, PYCHEOPS \citep{Maxted_2022}}

\appendix

\section{Summary of Observations}

The times, locations, telescopes, and observer names for the \planet \ observations analyzed in this work are given in Table \ref{table:observation table}. The time given is the length of the full observation but useful data may not have been acquired during that entire period.

\begin{deluxetable*}{ccccccc}[h!]
\tabletypesize{\scriptsize}
\tablecaption{Observations Summary \label{table:observation table}}
\tablehead{
  \colhead{eVscope ID} & 
  \colhead{eVscope} &
  \colhead{UTC Start} &
  \colhead{Start Time} & 
  \colhead{End Time}$^a$ & 
  \colhead{Observer} & 
  \colhead{Location} \\
  \colhead{(abbreviated)} &
  \colhead{Model} &
  \colhead{Date} &
  \colhead{(hh:mm:ss)} &
  \colhead{(hh:mm:ss)} &
  \colhead{} &
  \colhead{}}
  \startdata
257 & 1 & 2021 Nov 18 & 19:13:08 & 20:43:19  & Kuossari & Tervakoski, Finland \\ 
 ydh & 1 & 2021 Nov 18 & 22:41:19 & 23:11:10  & Stein & Lakewood, NJ, USA \\ 
 ka5 & 1 & 2021 Nov 18 & 23:12:51 & 23:16:53  & Mountz & Louisville, KY, USA \\ 
 rcn & 2 & 2021 Nov 18 & 23:14:21 & 03:46:22  & Randolph & Athens, GA, USA \\
 un8 & 1 & 2021 Nov 18 & 23:20:08 & 02:01:20  & Randolph & Athens, GA, USA\\ 
 24s & 1 & 2021 Nov 19 & 00:23:36 & 04:10:57  & Ostrem & Ames, IA, USA \\ 
 p6h & 1 & 2021 Nov 19 & 00:27:55 & 02:16:24  & Sivayogan & Raleigh, NC, USA \\ 
 sdp & 1 & 2021 Nov 19 & 01:01:20 & 05:12:44  & Wagner & Tucson, AZ, USA \\ 
 gau & 1 & 2021 Nov 19 & 01:35:23 & 04:47:26  & Loose & San Diego, CA, USA\\  
 p6r & 1 & 2021 Nov 19 & 04:58:17 & 05:23:01  & Dalba & Santa Cruz, CA, USA \\ 
 3pw & 1 & 2021 Nov 19 & 06:02:37 & 07:20:14  & Gamurot & Kapolei, HI, USA \\ 
 66d & 1 & 2021 Nov 19 & 08:27:40 & 12:03:20  & Tsuchiyama & Tokyo, Japan \\ 
 8cm & 1 & 2021 Nov 19 & 08:44:17 & 11:57:34  & Shimizu & Sayo-cho, Hyogo, Japan \\
 y6x & 1 & 2021 Nov 19 & 09:05:34 & 13:21:17  & Goto & Toyonaka, Osaka, Japan \\ 
 26e & 1 & 2021 Nov 19 & 09:47:06 & 12:25:55  & Fukui & Tsukuba, Ibaraki, Japan \\ 
 zqq & 1 & 2021 Nov 19 & 12:50:32 & 13:52:24  & Sharon & Singapore, Singapore \\
 jpg & 1 & 2021 Nov 19 & 16:58:59 & 18:32:02  & Simpson & Thirsk, UK\\  
 tuf & 1 & 2021 Nov 19 & 17:10:31 & 21:52:12  & Laugier & Simiane Collongue, France \\ 
 zh6 & 1 & 2021 Nov 19 & 17:22:46 & 18:55:27  & Mirwald & Munich, Germany\\ 
 p4t & 1 & 2021 Nov 19 & 17:35:37 & 21:29:06  & Carrera & Barcelona, Spain \\
 7bt & 1 & 2021 Nov 19 & 17:52:08 & 21:32:19  & Manganese & Cavaillon, France \\
 nsp & 1 & 2021 Nov 19 & 17:59:18 & 18:29:08  & Douezy & Serignan, France \\  
 5m4 & 1 & 2021 Nov 19 & 18:05:36 & 21:57:31  & Picard & Courthezon, France\\ 
 8zz & 1 & 2021 Nov 19 & 18:49:42 & 21:55:37  & Savonnet & Trevignin, France\\ 
 etx & 1 & 2021 Nov 19 & 19:24:08 & 22:19:38  & Guillet & Caen, France \\   
 ujx & 1 & 2021 Nov 19 & 22:16:59 & 01:03:12  & Simard & Mirabel, Québec, Canada \\ 
 2mv & 1 & 2021 Nov 19 & 22:45:55 & 02:52:39  & Parker & High Bridge, NJ, USA \\ 
 vwh & 1 & 2021 Nov 19 & 22:52:53 & 23:22:44  & Trudel & Saint-Lazare, Québec, Canada\\
 qer & 1 & 2021 Nov 19 & 23:06:02 & 03:12:28  & Will & Raleigh, NC, USA \\ 
 rp9 & 1 & 2021 Nov 19 & 23:12:10 & 02:24:10  & Runge  & Durham, NC, USA\\ 
 rcn-B & 2 & 2021 Nov 19 & 23:13:07 & 03:38:29  & Randolph & Athens, GA, USA \\ 
 un8-B & 1 & 2021 Nov 19 & 23:13:50 & 03:38:47  & Randolph & Athens, GA, USA\\ 
 rev & 1 & 2021 Nov 20 & 00:34:44 & 03:10:17  & Primm & Austin, TX, USA\\ 
 sdp-B & 1 & 2021 Nov 20 & 01:01:12 & 04:32:26  & Wagner & Tucson, AZ, USA\\ 
 gau-B & 1 & 2021 Nov 19 & 01:25:18 & 05:03:18  & Loose & San Diego, CA, USA \\ 
 p6r-B & 1 & 2021 Nov 19 & 01:41:35 & 05:28:08  & Dalba & Santa Cruz, CA, USA \\
 2n9 & 1 & 2021 Nov 20 & 03:06:56 & 05:33:22  & Esposito & Telegraph City, CA, USA\\ 
 fqe & 1 & 2021 Nov 20 & 04:01:11 & 05:41:29  & Esposito & Telegraph City, CA, USA\\
 33r & 1 & 2021 Nov 20 & 04:44:09 & 07:34:08  & Shih & Kahului, HI, USA \\  
 3pw-B & 1 & 2021 Nov 20 & 04:44:38 & 07:09:16  & Gamuro & Kapolei, HI, USA\\ 
 8cm-B & 1 & 2021 Nov 20 & 08:40:41 & 10:43:37  & Shimizu & Sayo-cho, Hyogo, Japan \\
 y6x-B & 1 & 2021 Nov 20 & 08:41:44 & 10:44:37  & Goto & Toyonaka, Osaka, Japan \\  
 \nodata & \nodata & 2021 Nov 19 & 01:27:41 & 05:41:08 & Lewin$^b$ & Glendora, CA, USA \\
 \nodata & \nodata & 2021 Nov 19 & 22:12:20 & 01:10:13 & Silvis$^b$ & Bourne, MA, USA \\
\enddata
\tablenotetext{a}{Some end times occur on the following day.}
\tablenotetext{b}{Citizen astronomers affiliated with Exoplanet Watch, not using eVscopes. See Section \ref{sec:EW_obs}.}
\end{deluxetable*}

\clearpage 

\bibliography{references}
\bibliographystyle{aasjournal}




\end{document}